\documentstyle[amsfonts,aps]{revtex}
\begin{document}
\title{Decoherence in closed and open systems}
\author{Mario Castagnino}
\address{CONICET-Instituto de Astronom\'{i}a y F\'{i}sica del Espacio\\
Casilla de Correos 67, Sucursal 28, 1428, Buenos Aires, Argentina}
\author{Roberto Laura}
\address{Facultad de Ciencias Exactas, Ingenier\'{i}a y Agrimensura, Universidad\\
Nacional de Rosario\\
Av. Pellegrini 250, 2000, Rosario, Argentina}
\author{Olimpia Lombardi}
\address{CONICET-Universidad Nacional de Quilmes\\
Rivadavia 2328, 6${{}^{o}}$ Derecha, 1034, Buenos Aires, Argentina}
\maketitle

\begin{abstract}
A generalized formal framework for decoherence, that can be used both in
open and closed quantum systems, is sketched. In this context, the
relationship between the decoherence of a closed system and the decoherence
of its subsystems is studied, and the corresponding decoherence times, $%
t_{DU}$ for the closed system and $t_{DS}$ for the open system, are defined:
for macroscopic systems, $t_{DU}\gg t_{DS}$. Finally, it is shown that the
application of the new formal framework to a well-known model leads to
physically adequate results.

PACS N${{}^o}$ 03.65.Yz, 03.65.Db
\end{abstract}

\section{Introduction}

The peculiar features of quantum mechanics are mainly due to the principle
of superposition and its consequence, the phenomenon of interference.
Therefore, any attempt to explain how classicality emerges from quantum
behavior must include two elements: a process through which interference
vanishes, and a resulting superselection rule that precludes superpositions.
Decoherence is the process that cancels interference and leads to the rule
that selects the candidates for classical states.

Historically, decoherence was conceived in terms of explaining how a
coherent pure state becomes a final stable decohered mixture with no
interference terms. In other words, the task was to explain how the state of
a quantum system goes from the frontier of the convex set of states to an
interior point of the convex. On this basis, decoherence was studied in
closed as in open systems. Schematically, three periods can be identified in
the development of this general program:

\begin{itemize}
\item  {\bf First period: Closed systems }(van Kampen \cite{van Kampen}, van
Hove \cite{van Hove}, Daneri {\it et al. }\cite{Daneri}). In order to
understand how classical macroscopic features arise from quantum microscopic
behavior, ''gross'' observables are defined, and the states that are
indistinguishable for a macroscopic observer are described by the same
coarse-grained state $\rho _{G}(t)$. When the evolution of $\rho _{G}(t)$
(or of the expectation value of the gross observables) is studied, it is
proved that $\rho _{G}(t)$ reaches equilibrium in a relaxation time $t_{R}$;
therefore, $\rho _{G}(t)$ decoheres in its own eigenbasis after a
decoherence time $t_{D}=t_{R}$.

This approach was rooted in the traditional study of irreversible processes.
The main problem of this period was the fact that decoherence times computed
with this primitive formalism turned out to be too long to account for
experimental data (see \cite{Omnes}).

\item  {\bf Second period: Open systems}. The open system $S$ is considered
in interaction with its environment $E$, and the evolution of the reduced
state $\rho _{S}(t)=Tr_{E}\rho (t)$ is studied. The so-called
''environment-induced decoherence (EID) approach'' (Zeh \cite{Zeh}, Zurek 
\cite{Zurek-1} \cite{Zurek-2}) proves that, since the states of $E$ become
rapidly orthogonal, the interference terms of $\rho _{S}(t)$ rapidly vanish
and $\rho _{S}(t)$ decoheres in an adequate pointer basis after a very short
decoherence time $t_{D}$; this result solves the main problem of the first
period.

This approach is rooted in the theory of quantum measurements, where the
open system $S$ interacts with a measurement apparatus $M$, and the
evolution correlates the states of both systems. In EID, $E$ plays the role
of $M$, and it is said that $E$ continuously measures $S.$ At present this
formalism has been applied to a wide range of models, and its results have
many experimental confirmations (see \cite{Ex}). However, the EID approach
still has to face three conceptual difficulties: (a) it cannot be applied to
closed systems, in particular, to the universe; according to Zurek, the
issue of the classicality of closed systems or of the universe as a whole
cannot even be posed (see \cite{Zurek-closed}, p.181); (b) it does not
supply a criterion for deciding where to place the cut between system and
environment; as Zurek himself admits, this is a serious problem for the
foundation of the whole EID program (see \cite{Zurek-cut}, p.22); (c) it
does not provide a simple general definition of the pointer basis (see \cite
{Zurek-pointer}).

\item  {\bf Third period: Closed and open systems. } Although at present EID
is still considered the ''orthodoxy'' in the subject (\cite{Bub}), other
approaches have been proposed to face its problems, in particular, the
closed-systems problem (Diosi, Milburn, Penrose, Casati and Chirikov, Adler 
\cite{Penrose}). Some of these methods are clearly ''non-dissipative'' (see 
\cite{O'Connel}), that is, not based on the dissipation of energy from the
system to the environment. Among them, we have developed the self-induced
decoherence (SID) approach, according to which a closed quantum system with
continuous spectrum may decohere by destructive interference and reach a
final state where the classical limit can be rigorously obtained (\cite{SID}%
).
\end{itemize}

In spite of the fact that the theories of decoherence in closed and open
systems coexist in the third period, in the literature both kinds of
approaches are usually conceived as alternative scenarios for decoherence,
or even as formalisms dealing with different phenomena\cite{Max}. In this
paper we will argue that this is not the case; on the contrary, formalisms
originally devised to deal just with closed or open systems can be subsumed
under a more general theoretical framework where the problems of both kinds
of approaches can be successfully faced. Our main aim is to present a new
conceptual perspective that will clarify some points that still remain
rather obscure in the literature on the subject. In particular, we will
develop our argument by comparing SID and EID; we will show that both
theories can be viewed as complementary and leading to compatible results.

On this basis, the paper is organized as follows. In Section II we will
present a general framework for decoherence and show how SID and EID can be
expressed in this theoretical context. Such a presentation will allow us to
explain, in Section III, the relationship between decoherence in closed and
open systems. Section IV is devoted to stress the compatibility between the
results obtained with SID\ and EID, in particular with respect to the
decoherence time. In Section V we will study in detail a well-known example,
showing how some results not correctly interpreted in previous works can be
understood from the new perspective. Finally, in Section VI we will draw our
conclusions.

\section{Observables, expectation values and weak limits}

As emphasized by Omn\'{e}s \cite{Omnes1}, decoherence is just a particular
case of the general problem of irreversibility in quantum mechanics. The
problem of irreversibility can be roughly expressed in the following terms.
Since the quantum state $\rho (t)$ follows an unitary evolution, it cannot
reach a final equilibrium state for $t\rightarrow \infty $. Therefore, if
the non-unitary evolution towards equilibrium is to be accounted for, a
further element has to be added to the unitary evolution. From the most
general viewpoint, this element consists in the splitting of the maximal
information about the system into a relevant part and an irrelevant part:
whereas the irrelevant part is disregarded, the relevant part is retained
and its evolution may reach a final equilibrium situation.

This broadly expressed idea can be rephrased in operators language. The
maximal information about the system is given by the set of all its
potentially possible observables. By selecting a particular subset ${\cal O}$
of this set, we restrict the maximal information to a relevant part: the
expectation values $\langle O_{R}\rangle _{\rho (t)}$ of the observables $%
O_{R}\in {\cal O}$ express the relevant information about the system. Of
course, the decision about which observables are to be considered as
relevant depends on the particular purposes in each situation; but without
this restriction, irreversible evolutions cannot be described.

Since decoherence is an irreversible process, it must include the splitting
of the whole set of observables into the relevant subset ${\cal O}$ and the
irrelevant subset. In fact, it is easy to see that the different approaches
to decoherence, when considered from this point of view, always select a set 
${\cal O}$ of relevant observables in terms of which the time behavior of
the system is described: gross observables in van Kampen \cite{van Kampen},
macroscopic observables of the apparatus in Daneri {\it et al.} \cite{Daneri}%
, observables of the open system in EID \cite{Zeh} \cite{Zurek-1} \cite
{Zurek-2}, relevant observables in Omn\'{e}s \cite{Omnes1}, van Hove
observables in SID \cite{SID}, etc.

Once the essential role played by the selection of the relevant observables
is clearly understood, the phenomenon of decoherence can be explained in
three general steps:

\begin{enumerate}
\item  {\bf First step:} The set ${\cal O}$ of relevant observables is
defined.

\item  {\bf Second step:} The expectation value $\langle O_{R}\rangle _{\rho
(t)}$, for any $O_{R}\in {\cal O}$, is obtained. This step can be formulated
in two different but equivalent ways:

\begin{itemize}
\item  $\langle O_{R}\rangle _{\rho (t)}$ is computed as the expectation
value of $O_{R}$ in the unitarily evolving state $\rho (t)$.

\item  A coarse-grained state $\rho _{G}(t)$ is defined by $\langle
O_{R}\rangle _{\rho (t)}=\langle O_{R}\rangle _{\rho _{G}(t)}$ for any $%
O_{R}\in {\cal O}$ (see Appendix A, eq.(\ref{A-4})), and its non-unitary
evolution (governed by a master equation) is computed.
\end{itemize}

\item  {\bf Third step:} It is proved that $\langle O_{R}\rangle _{\rho
(t)}=\langle O_{R}\rangle _{\rho _{G}(t)}$ reaches a final equilibrium value 
$\langle O_{R}\rangle _{\rho _{*}}$: 
\begin{equation}
\lim_{t\rightarrow \infty }\langle O_{R}\rangle _{\rho
(t)}=\lim_{t\rightarrow \infty }\langle O_{R}\rangle _{\rho _{G}(t)}=\langle
O_{R}\rangle _{\rho _{*}}  \label{2-0}
\end{equation}

This means that the coarse-grained state $\rho _{G}(t)$ evolves towards a
final equilibrium state (see Appendix A, eq.(\ref{A-6})): 
\begin{equation}
\lim_{t\rightarrow \infty }\langle O_{R}\rangle _{\rho _{G}(t)}=\langle
O_{R}\rangle _{\rho _{G*}}  \label{2-00}
\end{equation}

The final equilibrium state $\rho _{*}$ is obviously diagonal in its own
eigenbasis, which turns out to be the final pointer basis. But, as follows
from eq.(\ref{2-0}), the unitarily evolving quantum state $\rho (t)$ of the
whole system {\it has only a} {\it weak limit}: 
\begin{equation}
W-\lim_{t\rightarrow \infty }\rho (t)=\rho _{*}  \label{2-000}
\end{equation}
As a consequence, the coarse-grained state $\rho _{G}(t)$ also has a weak
limit, as follows from eq.(\ref{2-00}): 
\begin{equation}
W-\lim_{t\rightarrow \infty }\rho _{G}(t)=\rho _{G*}  \label{2-0000}
\end{equation}

These weak limits mean that, although the off-diagonal terms of $\rho (t)$
never vanish through the unitary evolution, the system decoheres {\it from
an observational point of view}, that is, from the viewpoint given by any
relevant observable $O_{R}\in {\cal O}$.
\end{enumerate}

From this general perspective, it turns out to be clear that decoherence is
a coarse-grained process that leads to classicality in a coarse-grained
sense. In fact, the phenomenon of interference is suppressed because the
off-diagonal terms of $\rho (t)$ and $\rho _{G}(t)$ vanish from the
viewpoint of the relevant observables, and the superselection rule that
precludes superpositions retains the states defined by the corresponding
pointer bases.

In the next subsections we will argue for the generality of this theoretical
framework by applying it to the SID and the EID approaches. This will show
that, in spite of the fact that SID deals with closed systems and EID
describes open systems, the general concept of decoherence expressed by
steps 1 to 3 lies behind both approaches.

\subsection{SID: decoherence in closed systems}

In the SID approach, the three steps are explicit in the formalism. For
conciseness, we will present the theory in the simplest case (for more
general cases, see \cite{SID}).

1. {\bf First step:} Let us consider a quantum system endowed with a
Hamiltonian $H$ with continuous spectrum: $H|\omega \rangle =\omega |\omega
\rangle $, $\omega \in [0,\infty )$. A generic observable reads 
\begin{equation}
O=\int_{0}^{\infty }\int_{0}^{\infty }\widetilde{O}(\omega ,\omega ^{\prime
})|\omega \rangle \langle \omega ^{\prime }|\,d\omega d\omega ^{\prime }
\label{2-1}
\end{equation}
where $\widetilde{O}(\omega ,\omega ^{\prime })$ is any kernel or
distribution. The restriction in the set of observables is introduced by
considering only the {\it van Hove operators}, whose components are given by 
\begin{equation}
\widetilde{O}_{R}(\omega ,\omega ^{\prime })=O(\omega )\delta (\omega
-\omega ^{\prime })+O(\omega ,\omega ^{\prime })  \label{2-2}
\end{equation}
where $O(\omega ,\omega ^{\prime })$ is a regular function. Then, the
relevant observables have the following form: 
\begin{equation}
O_{R}=\int_{0}^{\infty }O(\omega )|\omega )\,d\omega +\int_{0}^{\infty
}\int_{0}^{\infty }O(\omega ,\omega ^{\prime })|\omega ,\omega ^{\prime
})\,d\omega d\omega ^{\prime }  \label{2-3}
\end{equation}
where $|\omega )=|\omega \rangle \langle \omega |$ and $|\omega ,\omega
^{\prime })=|\omega \rangle \langle \omega ^{\prime }|$. These relevant
observables belong to the van Hove space ${\cal O}_{VH}$, whose basis is $%
\left\{ |\omega ),|\omega ,\omega ^{\prime })\right\} $.\footnote{%
This restriction on operators does not diminish the generality of SID, since
the observables not belonging to the van Hove space are not experimentally
accessible and, for this reason, in practice they are always approximated,
with the desired precision, by regular observables for which the approach
works satisfactorily (for a full argument, see \cite{Studies}).} States $%
\rho $ are represented by linear functionals on ${\cal O}_{VH}$, that is,
they belong to the dual space ${\cal O}_{VH}^{\prime }$ and read 
\begin{equation}
\rho =\int_{0}^{\infty }\rho (\omega )(\omega |\,d\omega +\int_{0}^{\infty
}\int_{0}^{\infty }\rho (\omega ,\omega ^{\prime })(\omega ,\omega ^{\prime
}|\,d\omega d\omega ^{\prime }  \label{2-4}
\end{equation}
where $\{(\omega |,(\omega ,\omega ^{\prime }|\}$ is the cobasis of $\left\{
|\omega ),|\omega ,\omega ^{\prime })\right\} $, that is, the basis of $%
{\cal O}_{VH}^{\prime }$. States must satisfy the usual requirements: $\rho
(\omega )$ is real and positive and $\int_{0}^{\infty }\rho (\omega )d\omega
=1$. We also require that $\rho (\omega ,\omega ^{\prime })$ be a regular
function. Under these conditions, states belong to a convex set $S\subset 
{\cal O}_{VH}^{\prime }$.

2. {\bf Second step:} The expectation value of the observable $O_{R}\in 
{\cal O}_{VH}$ in the state $\rho \in S$ can be computed as the action of
the functional $\rho $ on the operator $O_{R}$: 
\begin{equation}
\langle O_{R}\rangle _{\rho }=(\rho |O_{R})=\int_{0}^{\infty }\rho
^{*}(\omega )O(\omega )\,d\omega +\int_{0}^{\infty }\int_{0}^{\infty }\rho
^{*}(\omega ,\omega ^{\prime })O(\omega ,\omega ^{\prime })\,d\omega d\omega
^{\prime }  \label{2-5}
\end{equation}
where $\rho (\omega )$ and $O(\omega )$ are such that the first integral is
well defined. The time evolution of this expectation value is given by 
\begin{equation}
\langle O_{R}\rangle _{\rho (t)}=\int_{0}^{\infty }\rho ^{*}(\omega
)O(\omega )\,d\omega +\int_{0}^{\infty }\int_{0}^{\infty }\rho ^{*}(\omega
,\omega ^{\prime })O(\omega ,\omega ^{\prime })\,e^{i\frac{\omega -\omega
^{\prime }}{\hbar }t}\,d\omega d\omega ^{\prime }  \label{2-6}
\end{equation}

3. {\bf Third step:} Since the function $\rho ^{*}(\omega ,\omega ^{\prime
})O(\omega ,\omega ^{\prime })$ is regular (precisely, it is ${\Bbb L}_{1}$
in variable $\nu =\omega -\omega ^{\prime }$), the Riemann-Lebesgue theorem
can be applied to eq.(\ref{2-6}).\footnote{%
The Riemann-Lebesgue theorem, which mathematically expresses the phenomenon
of {\it destructive interference}, states that, if $f(\nu )\in {\Bbb L}_{1}$%
, 
\[
\lim_{t\rightarrow \infty }\int d\nu f(\nu )e^{i\nu t}=0 
\]
} As a consequence, the second term vanishes: 
\begin{equation}
\lim_{t\rightarrow \infty }\langle O_{R}\rangle _{\rho (t)}=\int_{0}^{\infty
}\rho ^{*}(\omega )O(\omega )\,d\omega  \label{2-6'}
\end{equation}
This means that, for $t\rightarrow \infty $, the expectation value of any
observable $O_{R}\in {\cal O}_{VH}$ in the state $\rho \in S$ can be
computed as if the system were in a final stable state $\rho _{*}$: 
\begin{equation}
\lim_{t\rightarrow \infty }\langle O_{R}\rangle _{\rho (t)}=\langle
O_{R}\rangle _{\rho _{*}}  \label{2-7}
\end{equation}
where $\rho _{*}=\int_{0}^{\infty }\rho ^{*}(\omega )(\omega |\,d\omega $
has only singular diagonal terms in the eigenbasis of the Hamiltonian. This
result can also be expressed as a weak limit: 
\begin{equation}
W-\lim_{t\rightarrow \infty }\rho (t)=\rho _{*}  \label{2-8}
\end{equation}
This means that the system decoheres in the eigenbasis of the Hamiltonian;
these states are stationary and, therefore, completely robust.

Summing up, through steps 1 to 3 (see the coincidence between eqs.(\ref{2-7}%
)-(\ref{2-8}) and eqs.(\ref{2-0})-(\ref{2-000})), SID cancels interference
and leads to the superselection rule that precludes superpositions.

More general models have been treated with the SID approach \cite{SID}, and
decoherence times have been computed \cite{DT}; the foundations of the
theory have also been conceptually explained \cite{Studies}. Although SID
strictly applies in the continuous case, it also leads to approximate
decoherence in quasi-continuous models, that is, discrete models where (i)
the energy spectrum is quasi-continuous, i.e., has a small discrete energy
spacing, and (ii) the functions of energy used in the formalism are such
that the sums in which they are involved can be approximated by Riemann
integrals. These conditions are applied to a concrete example in \cite
{Alimañas} where it is shown that, in spite of the fact that, strictly
speaking, a system with discrete spectrum never reaches equilibrium due to
Poincar\'{e} recurrence, for times $t\ll t_{R}$, where $t_{R}$ is the
recurrence time, the discrete spectrum can be approximated by a continuous
spectrum when the involved functions satisfy the usual conditions of
regularity and integrability. These conditions are rather weak: in fact, the
overwhelming majority of the physical models studied in the literature on
dynamics, thermodynamics, quantum mechanics and quantum field theory are
quasi-continuous, and the well-known strategy for transforming sums in
integrals is applied.\footnote{%
If we have the sum: 
\[
\frac{2\pi \hbar }{L}\sum_{0}^{p}f_{p} 
\]
where $L$ is ''the size of the box'', from the uncertainty principle we can
make $\frac{2\pi \hbar }{L}\simeq \Delta p$ and, therefore, the strategy is: 
\[
\sum_{0}^{p}f_{p}\Delta p\longrightarrow \int_{0}^{p}f(p)\,dp 
\]
}

\subsection{EID: decoherence in open systems}

In the case of the EID approach, steps 1 to 3 are usually not explicit in
the formalism. However, the theory can be rephrased in such a way that it
can be analyzed from the general framework introduced at the beginning of
this section.

1. {\bf First step:} Let us consider a closed system $U$ that can be
decomposed into a proper system $S$ and its environment $E$. Let ${\cal H}=%
{\cal H}_{S}\otimes $ ${\cal H}_{E}$, where ${\cal H}_{S}$ is the Hilbert
space of $S$ and ${\cal H}_{E}$ the Hilbert space of $E$. The corresponding
von Neumann-Liouville space of $U$ is ${\cal L}={\cal H\otimes H=L}%
_{S}\otimes $ ${\cal L}_{E}$, where ${\cal L}_{S}={\cal H}_{S}\otimes $ $%
{\cal H}_{S}$ and ${\cal L}_{E}={\cal H}_{E}\otimes $ ${\cal H}_{E}$. A
generic observable belonging to ${\cal L}$ reads 
\begin{equation}
O=O_{S}\otimes O_{E}\text{ }\in {\cal L}\text{, \quad with }O_{S}\in {\cal L}%
_{S}\text{ and }O_{E}\in {\cal L}_{E}  \label{2-9}
\end{equation}
e.g., with coordinates $(O_{i\alpha j\beta })=(O_{ij}O_{\alpha \beta })$,
where $i,j,...$ are the indices corresponding to ${\cal H}_{S}$, and $\alpha
,\beta ,...$ are the indices corresponding to ${\cal H}_{E}$. The relevant
observables are those having the following form:

\begin{equation}
O_{R}=O_{S}\otimes I_{E}\in {\cal O}_{R}\text{, \quad with coordinates }%
(O_{ij}\delta _{\alpha \beta })  \label{2-10}
\end{equation}
where $I_{E}$ is the identity operator in ${\cal L}_{E}$. Therefore, ${\cal O%
}_{R}\subset {\cal L}$ is the subset of the relevant observables, that is,
those corresponding to the proper system $S$.

2. {\bf Second step:} The expectation value of any observable $O_{R}\in $ $%
{\cal O}_{R}$ in the state $\rho $ of $U$ reads

\begin{equation}
\langle O_{R}\rangle _{\rho }=Tr\,(\rho O_{R})=\sum_{ij\alpha \beta }\rho
_{i\alpha j\beta }^{*}\,O_{ij}\,\delta _{\alpha \beta
}=\sum_{ij}O_{ij}\sum_{\alpha \beta }\rho _{i\alpha j\beta }^{*}\,\delta
_{\alpha \beta }=\sum_{ij}O_{ij}\sum_{\alpha }\rho _{i\alpha j\alpha }^{*}
\label{2-11}
\end{equation}
If we define the reduced density operator $\rho _{S}$ by tracing over the
environmental degrees of freedom, we obtain

\begin{equation}
\rho _{S}=Tr_{E}\,\rho \in {\cal L}_{S}^{\prime }\text{, \quad with
coordinates }\left( \sum_{\alpha }\rho _{i\alpha j\alpha }\right) =(\rho
_{ij})  \label{2-12}
\end{equation}
where ${\cal L}_{S}^{\prime }$ is the dual space of ${\cal L}_{S}$.
Therefore, the expectation value $\langle O_{R}\rangle _{\rho (t)}$ can be
expressed as 
\begin{equation}
\langle O_{R}\rangle _{\rho (t)}=Tr\,\left( \rho (t)\,O_{R}\right)
=Tr\,\left( \rho (t)(O_{S}\otimes I_{E})\right) =Tr\left( \rho
_{S}(t)\,O_{S}\right) =\langle O_{S}\rangle _{\rho _{S}(t)}  \label{2-13}
\end{equation}

3. {\bf Third step:} The EID approach studies the time evolution of the
reduced density operator $\rho _{S}(t)$ governed by an effective master
equation. For many physical models where the space ${\cal O}_{R}$ has a
finite number of dimensions, this approach shows that, for $t\rightarrow
\infty $, $\rho _{S}(t)$ strongly reaches an equilibrium state $\rho _{S*}$
(see Appendix A, eq.(\ref{A-11})):

\begin{equation}
\rho _{S}(t)\longrightarrow \rho _{S*}  \label{2-14}
\end{equation}
Since $\rho _{S*}$ is obviously diagonal in its eigenbasis, the system $S$
decoheres in the eigenbasis of $\rho _{S*}$, which turns out to be the final
pointer basis. But if we take into account the definition of $\rho _{S}$ as
a partial trace (see eq.(\ref{2-12})), we can obtain the limit of the
expectation values of eq.(\ref{2-13}) as 
\begin{equation}
\lim_{t\rightarrow \infty }\langle O_{S}\rangle _{\rho
_{S}(t)}=\lim_{t\rightarrow \infty }\langle O_{R}\rangle _{\rho (t)}=\langle
O_{S}\rangle _{\rho _{S_{*}}}=\langle O_{R}\rangle _{\rho _{*}}  \label{2-15}
\end{equation}
where $\rho _{*}$ is such that $\rho _{S*}$ results from the projection of $%
\rho _{*}$ onto ${\cal O}_{R}$ (see Appendix A). Therefore, for any
observable $O_{R}\subset $ ${\cal O}_{R}$, 
\begin{equation}
\lim_{t\rightarrow \infty }\langle O_{R}\rangle _{\rho (t)}=\langle
O_{R}\rangle _{\rho _{*}}  \label{2-16}
\end{equation}
This result can also be expressed as a weak limit: 
\begin{equation}
W-\lim_{t\rightarrow \infty }\rho (t)=\rho _{*}  \label{2-17}
\end{equation}

Summing up, through steps 1 to 3, EID also suppresses interference and leads
to the superposition rule that precludes superpositions.

If the just obtained eqs.(\ref{2-16}) and (\ref{2-17}) are compared with the
corresponding eqs.(\ref{2-7}) and (\ref{2-8}) in the SID approach, the
similarity between them can be easily verified. This shows that the EID
approach can also be formulated from the viewpoint of the closed composite
system $U$ and, from this perspective, it can be explained in the context of
the general framework introduced at the beginning of this section. In other
words, the splitting of the closed system into a proper open system and an
environment is just a way of selecting the relevant observables of the
closed system. In fact, the environment may be external -such as particles
if air of photons scattered of the system- or internal -such as collections
of phonons or other internal excitations-; thus, the splitting of $U$
consists in a decision about which degrees of freedom are of direct interest
to the observer and which are irrelevant. Since the same system $U$ can be
decomposed in many different ways, there is nothing essential in such a
decomposition: there is no need of an unequivocal criterion for placing the
cut between ''the'' system and ''the'' environment. From this perspective,
the essential physical fact is that, among all the possible decompositions
of a closed system, there are some that lead to identify a subset of
relevant observables for which the system decoheres.

The EID approach is usually applied to models with discrete energy spectrum.
However, in most cases the sums introduced by the formalism are replaced by
Riemann integrals because of the quasi-continuous character of the model
under study (see, e.g., eq.(3-11) of \cite{Paz-Zurek}). Nevertheless, there
seems to be particular examples where the EID formalism can be successfully
applied for times $t\ll t_{R}$, where $t_{R}$ is the recurrence time, in
spite of the fact that the conditions for quasi-continuity are not satisfied
(see \cite{Max}).

\section{Closed and open systems}

Since we have showed that decoherence in open and closed systems can be
understood in the context of a common general framework, now the
relationship between both cases can be studied. In particular, we will
explore under what conditions (i) the decoherence of a closed composite
system implies the decoherence of any of its open subsystems, and (ii) the
decoherence of the open subsystems implies the decoherence of the closed
composite system. The results obtained from this analysis will point to the
fact that the formalisms of decoherence for closed and open systems are
complementary, and both cooperate in the understanding of the same physical
phenomenon.

In order to develop our arguments, we will study the case of systems with
discrete and finite spectra; the obtained results can be extended to the
case of continuous spectrum under the assumption of the usual conditions of
quasi-continuity. This strategy does not involve a loss of physical
generality for the following reason. Theoretical results must always be
tested by numerical simulations; then, continuous functions need to be
approximated by discrete functions, and the numerical experiments are
performed for a progressively increasing number $N$ of degrees of freedom in
order to simulate the continuous situation. In the case of testing
decoherence results, such a procedure is completely reasonable from a
physical point of view. In fact, let us consider a closed system $U$
partitioned into two subsystems $S_{1}$ and $S_{2}$ such that $U=S_{1}\cup
S_{2}$. Let us call ${\cal O}_{0}$ the set of the discrete observables of $U$%
, and ${\cal O}_{1}$ and ${\cal O}_{2}$ the sets of discrete observables $%
O_{R1}$ and $O_{R2}$ that are relevant from the viewpoint of $S_{1}$ and $%
S_{2}$ respectively: 
\begin{eqnarray}
O_{R1} &=&O_{S1}\otimes I_{S2}\in {\cal O}_{1}\subset {\cal O}_{0}
\label{3-1} \\
O_{R2} &=&O_{S2}\otimes I_{S1}\in {\cal O}_{2}\subset {\cal O}_{0}
\label{3-2}
\end{eqnarray}
When $N$ is large enough, and under the conditions of quasi-continuity, the $%
U$-relevant observables belonging to the van Hove space ${\cal O}_{VH}$ can
be approximated by discrete observables belonging to ${\cal O}_{0}$ because
any distribution can always be approximated, in the context of integration,
by a discrete function with the desired precision. As a consequence,
numerical simulations will show that, if the system $U$ decoheres according
to SID, it approximately decoheres for any discrete observable $O_{U}\in 
{\cal O}_{0}$. Therefore, the assumption that the $S_{1}$-relevant
observables and the $S_{2}$-relevant observables belong to the set of the $U$%
-relevant observables does not diminish the physical generality of our
arguments.

\subsection{From the closed system to its open subsystems}

Let us consider the closed system $U$ partitioned into the two subsystems $%
S_{1}$ and $S_{2}$ (each subsystem can be thought as the environment of the
other). If ${\cal H}$ is the Hilbert space of $U$ with non-prime dimension $%
N=nm$, with $n,m\in {\Bbb N}$, ${\cal H}$ can be decomposed as 
\begin{equation}
{\cal H=H}_{1}\otimes {\cal H}_{2}  \label{3-3}
\end{equation}
where ${\cal H}_{1}$ and ${\cal H}_{2}$ are the Hilbert spaces of $S_{1}$
and $S_{2}$ respectively. Let $\{|i\rangle \}$ be a basis of ${\cal H}_{1}$ (%
$i=1,...,n$), and $\{|\alpha \rangle \}$ be a basis of ${\cal H}_{2}$ ($%
\alpha =1,...,m$). Then, a basis of ${\cal H}$ is 
\begin{equation}
\left\{ |i,\alpha \rangle \right\} =\left\{ |i\rangle \otimes |\alpha
\rangle \right\}  \label{3-4}
\end{equation}
The coordinates of the observables $O_{U}\in {\cal O}_{0}$, $O_{R1}\in {\cal %
O}_{1}$ and $O_{R2}\in {\cal O}_{2}$ are 
\begin{eqnarray}
(O_{i\alpha j\beta })\text{{}} &:&\text{ coordinates of }O_{U}  \label{3-4'}
\\
(O_{i\alpha j\beta }^{R1} &=&O_{ij}^{S1}\delta _{\alpha \beta })\text{{}{}}:%
\text{ coordinates of }O_{R1}\qquad (O_{ij}^{S1})\text{: coordinates of }%
O_{S1}\text{ }  \label{3-5} \\
(O_{i\alpha j\beta }^{R2} &=&\delta _{ij}O_{\alpha \beta }^{S2})\text{{}{}}:%
\text{ coordinates of }O_{R2}\qquad (O_{\alpha \beta }^{S2})\text{:
coordinates of }O_{S2}  \label{3-6}
\end{eqnarray}

If we assume that the closed system $U$ decoheres, the expectation value of
any observable $O_{U}\in {\cal O}_{0}$ in the state $\rho _{U}(t)$ of $U$
reaches a final stable value (see eqs.(\ref{2-0}) and (\ref{2-7})): 
\begin{equation}
\lim_{t\rightarrow \infty }\langle O_{U}\rangle _{\rho _{U}(t)}=\langle
O_{U}\rangle _{\rho _{U*}}  \label{3-7}
\end{equation}
This means that the state $\rho _{U}(t)$ has a weak limit (see eqs.(\ref
{2-000}) and (\ref{2-8})): 
\begin{equation}
W-\lim_{t\rightarrow \infty }\rho _{U}(t)=\rho _{U*}  \label{3-8}
\end{equation}
and, as a consequence, each one of the components $(\rho _{i\alpha j\beta
}(t))$ of $\rho _{U}(t)$ reaches a final stable value $(\rho _{*i\alpha
j\beta })$; these values are the coordinates of $\rho _{U*}$: 
\begin{equation}
(\rho _{i\alpha j\beta }(t))\longrightarrow (\rho _{*i\alpha j\beta })
\label{3-9}
\end{equation}
In particular, 
\begin{eqnarray}
(\rho _{i\alpha j\alpha }(t)) &\longrightarrow &(\rho _{*i\alpha j\alpha })
\label{3-10} \\
(\rho _{i\alpha i\beta }(t)) &\longrightarrow &(\rho _{*i\alpha i\beta })
\label{3-11}
\end{eqnarray}

By means of eq.(\ref{2-12}), let us now define the reduced density operators 
$\rho _{S1}$ and $\rho _{S2}$ by tracing over the degrees of freedom
corresponding to $S_{2}$ and $S_{1}$ respectively: 
\begin{eqnarray}
\rho _{S1} &=&Tr_{S2}\,\rho _{U}\text{, \quad with coordinates }\left(
\sum_{\alpha }\rho _{i\alpha j\alpha }=\rho _{ij}\right)  \label{3-12} \\
\rho _{S2} &=&Tr_{S1}\,\rho _{U}\text{, \quad with coordinates }\left(
\sum_{i}\rho _{i\alpha i\beta }=\rho _{\alpha \beta }\right)  \label{3-13}
\end{eqnarray}
From eqs.(\ref{3-10}) and (\ref{3-11}) we know that these coordinates also
reach their final equilibrium values: 
\begin{eqnarray}
\left( \rho _{ij}(t)=\sum_{\alpha }\rho _{i\alpha j\alpha }(t)\right)
&\longrightarrow &\left( \rho _{*ij}\right)  \label{3-14} \\
\left( \rho _{\alpha \beta }(t)=\sum_{i}\rho _{i\alpha i\beta }(t)\right)
&\longrightarrow &\left( \rho _{*\alpha \beta }\right)  \label{3-15}
\end{eqnarray}
where $(\rho _{*ij})$ and $(\rho _{*\alpha \beta })$ can be conceived as the
components of the final states $\rho _{S1*}$ and $\rho _{S2*}$ of $S_{1}$
and $S_{2}$ to which the reduced states $\rho _{S1}$ and $\rho _{S2}$ tend
respectively for $t\rightarrow \infty $ (see eq.(\ref{2-14})). Therefore,
the subsystems $S_{1}$ and $S_{2}$ also decohere for their respective
relevant observables (see eq.(\ref{2-15})): 
\begin{eqnarray}
\lim_{t\rightarrow \infty }\langle O_{R1}\rangle _{\rho _{U}(t)}
&=&\lim_{t\rightarrow \infty }\langle O_{S1}\rangle _{\rho _{S1}(t)}=\langle
O_{S1}\rangle _{\rho _{S1_{*}}}  \label{3-16} \\
\lim_{t\rightarrow \infty }\langle O_{R2}\rangle _{\rho _{U}(t)}
&=&\lim_{t\rightarrow \infty }\langle O_{S2}\rangle _{\rho _{S2}(t)}=\langle
O_{S2}\rangle _{\rho _{S2_{*}}}  \label{3-17}
\end{eqnarray}

Let us note that the argument does not depend on the particular partition of 
$U$ into $S_{1}$ and $S_{2}$. This means that, when the whole composite
system decoheres according to SID, the subsystems will also decohere no
matter how many degrees of freedom they have. In fact, if $U$ is a system of 
$N$ interacting oscillators, we can decide to split it into a single
oscillator as $S_{1}$ and the remaining $N-1$ oscillators as $S_{2}$: if $U$
decoheres, $S_{1}$ and $S_{2}$ also decohere. This conclusion shows that it
is {\it not always necessary} for the decoherence of an open system its
interaction with an environment with many, potentially infinite, degrees of
freedom: the decoherence of the whole composite system imposes a physical
situation as strong as to lead to the decoherence of any of its subsystems.

\subsection{From the open subsystems to the closed system}

Let us consider again the closed system $U$ partitioned into the subsystems $%
S_{1}$ and $S_{2}$ such that $U=S_{1}\cup S_{2}$, and whose Hilbert space $%
{\cal H}$ can be decomposed as 
\begin{equation}
{\cal H=H}_{1}\otimes {\cal H}_{2}  \label{3-18}
\end{equation}
where ${\cal H}_{1}$ and ${\cal H}_{2}$ are the Hilbert spaces of $S_{1}$
and $S_{2}$ respectively. If $\{|i\rangle \}$ is a basis of ${\cal H}_{1}$ ($%
i=1,...,n$) and $\{|\alpha \rangle \}$ is a basis of ${\cal H}_{2}$ ($\alpha
=1,...,m$), then a basis of ${\cal H}$ is 
\begin{equation}
\{|{\cal A}\rangle \}=\left\{ |i,\alpha \rangle \right\} =\left\{ |i\rangle
\otimes |\alpha \rangle \right\}  \label{3-19}
\end{equation}
For a different partition $U=S_{1}^{\prime }\cup S_{2}^{\prime }$, the
Hilbert space ${\cal H}$ of dimension $N=nm=kl$, with $n,m,k,l\in {\Bbb N}$,
can be decomposed as 
\begin{equation}
{\cal H=H}_{1}^{\prime }\otimes {\cal H}_{2}^{\prime }  \label{3-20}
\end{equation}
where ${\cal H}_{1}^{\prime }$ and ${\cal H}_{2}^{\prime }$ are the Hilbert
spaces of $S_{1}^{\prime }$ and $S_{2}^{\prime }$ respectively. If $%
\{|I\rangle \}$ is a basis of ${\cal H}_{1}^{\prime }$ ($I=1,...,k$) and $%
\{|\Gamma \rangle \}$ is a basis of ${\cal H}_{2}^{\prime }$ ($\Gamma
=1,...,l$), then another basis of ${\cal H}$ will be 
\begin{equation}
\{|{\cal A}^{\prime }\rangle \}=\left\{ |I,\Gamma \rangle \right\} =\left\{
|I\rangle \otimes |\Gamma \rangle \right\}  \label{3-21}
\end{equation}
The change of basis from $\{|{\cal A}\rangle \}$ to $\{|{\cal A}^{\prime
}\rangle \}$ can be performed by a linear transformation $A_{{\cal A}}^{%
{\cal A}^{\prime }}=A_{i}^{i^{\prime }}A_{{\cal \alpha }}^{{\cal \alpha }%
^{\prime }}$ such that 
\begin{equation}
|{\cal A}^{\prime }\rangle {\cal =}\sum_{{\cal A}}A_{{\cal A}}^{{\cal A}%
^{\prime }}|{\cal A}\rangle  \label{3-22}
\end{equation}

If we want to prove that the closed composite system $U$ decoheres, we have
to find the $(nm)^{2}-1$ real numbers that define the $(nm)^{2}$ complex
coordinates $(\rho _{*i\alpha j\beta })$ of the final state $\rho _{U*}$ of $%
U$.\footnote{%
A state represented by an $N\times N$ self-adjoint matrix has $N-1$ real
diagonal independent coordinates and $\frac{1}{2}(N^{2}-N)$ complex
off-diagonal coordinates. Therefore, the state is completely determined by $%
N-1+2\left[ \frac{1}{2}(N^{2}-N)\right] =N^{2}-1$ real numbers.} Let us
assume that the open subsystems $S_{1}$ and $S_{2}$ respectively decohere
for their relevant observables $O_{R1}\in {\cal O}_{1}$ and $O_{R2}\in {\cal %
O}_{2}$ (or $O_{S1}$ and $O_{S2}$) given by eqs.(\ref{3-1}) and (\ref{3-2}),
and whose corresponding coordinates are given by eqs.(\ref{3-5}) and (\ref
{3-6}). The decoherence of the subsystems means that (see eqs.(\ref{3-16})
and (\ref{3-17})) 
\begin{eqnarray}
\lim_{t\rightarrow \infty }\langle O_{R1}\rangle _{\rho _{U}(t)} &=&\langle
O_{R1}\rangle _{\rho _{U_{*}}}=\lim_{t\rightarrow \infty }\langle
O_{S1}\rangle _{\rho _{S1}(t)}=\langle O_{S1}\rangle _{\rho _{S1_{*}}}
\label{3-23} \\
\lim_{t\rightarrow \infty }\langle O_{R2}\rangle _{\rho _{U}(t)} &=&\langle
O_{R2}\rangle _{\rho _{U_{*}}}=\lim_{t\rightarrow \infty }\langle
O_{S2}\rangle _{\rho _{S2}(t)}=\langle O_{S2}\rangle _{\rho _{S2_{*}}}
\label{3-24}
\end{eqnarray}
where $\rho _{S1}(t)$ and $\rho _{S2}(t)$ are the reduced density operators
defined in eqs.(\ref{3-12}) and (\ref{3-13}), and $\rho _{S1*}$ and $\rho
_{S2*}$ are the final states to which the reduced states $\rho _{S1}(t)$ and 
$\rho _{S2}(t)$ tend respectively for $t\rightarrow \infty $ (see eq.(\ref
{2-14})). Therefore, 
\begin{eqnarray}
\sum_{\alpha }\rho _{*i\alpha j\alpha }\,O_{i\alpha j\alpha }^{R1} &=&\rho
_{*ij}\,O_{ij}^{S1}  \label{3-25} \\
\sum_{i}\rho _{*i\alpha i\beta }\,O_{i\alpha i\beta }^{R2} &=&\rho _{*\alpha
\beta }\,O_{\alpha \beta }^{S2}  \label{3-26}
\end{eqnarray}
Eqs.(\ref{3-25}) and (\ref{3-26}) represent a system of $(n^{2}-1)+(m^{2}-1)$
real equations which, in general, is not enough to lead to the $(nm)^{2}-1$
real numbers required to define the $(nm)^{2}$ coordinates $(\rho _{*i\alpha
j\beta })$ of $\rho _{U*}$. Nevertheless, we can introduce a different
partition of $U$: $U=S_{1}^{\prime }\cup S_{2}^{\prime }$. If we repeat the
argument for this new partition, we will obtain

\begin{eqnarray}
\sum_{\Gamma }\rho _{*I\Gamma J\Gamma }\,O_{I\Gamma J\Gamma }^{R^{\prime }1}
&=&\rho _{*IJ}\,O_{IJ}^{S^{\prime }1}  \label{3-27} \\
\sum_{I}\rho _{*I\Gamma I\Delta }\,O_{I\Gamma I\Delta }^{R^{\prime }2}
&=&\rho _{*\Gamma \Delta }\,O_{\Gamma \Delta }^{S^{\prime }2}  \label{3-28}
\end{eqnarray}
which represent a system of $(k^{2}-1)+(l^{2}-1)$ real equations. Eqs.(\ref
{3-27}) and (\ref{3-28}) are independent of eqs.(\ref{3-25}) and (\ref{3-26}%
) because the coefficients of the first equations are the $O_{IJ}^{S^{\prime
}1}$, $O_{\Gamma \Delta }^{S^{\prime }2}$, the coefficients of the last
equations are the $O_{ij}^{S1}$, $O_{\alpha \beta }^{S2}$, and these groups
of coefficients are not related by a linear transformation. Therefore, now
we have obtained $n^{2}+m^{2}+k^{2}+l^{2}-4$ real equations: if this number
of equations is enough to obtain the $(nm)^{2}-1$ real numbers required to
define the coordinates of $\rho _{U*}$, then we have proved the decoherence
of the whole system $U$ for the observables belonging to ${\cal O}_{1}\cup 
{\cal O}_{2}\cup {\cal O}_{1}^{\prime }\cup {\cal O}_{2}^{\prime }\subset 
{\cal O}_{0}$. If the $n^{2}+m^{2}+k^{2}+l^{2}-4$ real equations are not yet
sufficient to define $\rho _{U*}$, we can introduce further partitions up to
reach the necessary number of equations.\footnote{%
Let us consider the $n,m$ partitions corresponding to different rotated
bases \{$|{\cal A}^{\prime }{\cal \rangle }$ \}. In this case, the necessary
number of partitions is:
\par
\[
N=\frac{(nm)^{2}-1}{n^{2}+m^{2}-2}\cong \left( \frac{1}{n^{2}}+\frac{1}{m^{2}%
}\right) ^{-1} 
\]
when $m\gg 1$. If we keep $n$ finite and make $m\rightarrow \infty $, we
just need a finite number of partitions $N=n^{2}$.}

This argument can be used to study well-known models. For instance, let us
consider a set of $N$ oscillators, each one with its corresponding Hilbert
space ${\cal H}_{i}$ and the creation and annihilation operators $%
a_{i}^{\dagger }$, $a_{i}$; these operators define the bases of the ${\cal H}%
_{i}$ from a common vacuum $|0\rangle $. In this case, the complete Hilbert
space is ${\cal H=}\bigotimes_{i}{\cal H}_{i}$. But if we make a Bogoliubov
transformation: 
\begin{equation}
a_{i}^{\prime }=\sum_{i}\alpha _{i}^{j}a_{j}+\beta _{i}^{j}a_{j}^{\dagger
},\qquad a_{i}^{\prime \dagger }=\sum_{i}\alpha _{i}^{j*}a_{j}^{\dagger
}+\beta _{i}^{j*}a_{j}  \label{3-29}
\end{equation}
the new operators $a_{i}^{\prime \dagger }$, $a_{i}^{\prime }$ define the
corresponding bases of the new Hilbert spaces ${\cal H}_{i}^{\prime }$ and,
therefore, they introduce a different partition ${\cal H=}\bigotimes_{i}%
{\cal H}_{i}^{\prime }$ of the whole composite system.\footnote{%
If $\beta _{i}^{j}=0$, we remain in the same decomposition since $%
a_{i}^{\prime }=\sum_{i}\alpha _{i}^{j}a_{j}$ produces just a change of
basis in ${\cal H}_{i}$.} In this physical example, the argument developed
in this subsection can be easily applied.

Summing up, this argument shows that the decoherence of the subsystems of a
closed system is not a phenomenon of a different nature than or independent
of the decoherence of the whole composite system; on the contrary, there is
a close relationship between both phenomena. Therefore, the decoherence of
the whole closed system can be understood by studying the behavior of its
subsystems.

\section{Decoherence times}

In the previous sections we have showed how the SID and the EID approaches
to decoherence can be understood from a general theoretical framework, and
we have argued that there is a close link between the decoherence of a whole
closed system and the decoherence of its subsystems. Clearly, if this is the
case, there must be a meaningful relationship between the decoherence times
of the whole system and of its subsystems. This is the point that we will
address in this section.

\subsection{EID: decoherence time in open systems}

In several models studied by the EID approach, the decoherence time $t_{DS}$
of an open subsystem $S$ in interaction with its environment $E$ turns out
to be the relaxation time $t_{R}$ of the whole system $S\cup E$ multiplied
by a macroscopicity coefficient. For instance, in eq.(47) of \cite{Paz-Zurek}
or in eq.(3.136) of \cite{Ex}, 
\begin{equation}
t_{DS}=\left( \frac{\lambda _{DB}}{L_{0}}\right) ^{2}t_{R}  \label{4-1}
\end{equation}
where $\lambda _{DB}$ is the de Broglie length and $L_{0}$ is a macroscopic
characteristic length. In turn, in page 51 of \cite{Paz-Zurek}, 
\begin{equation}
t_{DS}=\left( \frac{\Delta x}{2L_{0}}\right) ^{2}t_{R}  \label{4-2}
\end{equation}
where $\frac{\Delta x}{2L_{0}}$ is the ratio between a microscopic and a
macroscopic characteristic lengths. In any case, $t_{D}$ is extremely short
since the macroscopicity ratios $\frac{\lambda _{DB}}{L_{0}}$ or $\frac{%
\Delta x}{2L_{0}}$ are extremely small (e.g. $10^{-20}$, see \cite{Paz-Zurek}%
). Therefore, $t_{D}\ll $ $t_{R}$.

\subsection{SID: decoherence time in closed systems}

In paper \cite{DT}, we have computed the decoherence time $t_{DU}$ of a
closed system $U$ in terms of the poles of the Hamiltonian resolvent and of
the initial conditions; in particular, we have shown that, if the
Hamiltonian and the initial conditions are trivial -that is, with just real
poles-, the decoherence time is infinite. In the Appendix B of that paper,
we have applied the method to a two-times evolution (application that can be
easily generalized to a n-times evolution). In this subsection we will
rephrase that appendix to show that, for an adequate choice of the
interactions, the characteristic times of the two-times evolution are the
decoherence time $t_{DS}$ of the open proper system and the decoherence time 
$t_{DU}$ of the closed composite system.

Let us consider a closed system $U$ partitioned into an open system $S$ and
its environment $E$, whose total Hamiltonian reads

\begin{equation}
H=H_{0}+V=H_{0}+\int_{0}^{\infty }\int_{0}^{\infty }[V^{(1)}(\omega ,\omega
^{\prime })+V^{(2)}(\omega ,\omega ^{\prime })]\,|\omega \rangle \langle
\omega ^{\prime }|\,d\omega d\omega ^{\prime }  \label{4-3}
\end{equation}
where $H_{0}$ is the free trivial Hamiltonian of $U$, $\{|\omega \rangle \}$
is its eigenbasis, $V^{(1)}$ represents the interaction between $S$ and $E$,
and $V^{(2)}$ represents the interaction of the parts of the environment $E$
among themselves.\footnote{%
Eq. (\ref{4-3}) is written with integrals for simplicity, but it could also
be expressed with sums in a discrete case.} We will also assume that $%
V^{(1)}(\omega ,\omega ^{\prime })\gg V^{(2)}(\omega ,\omega ^{\prime })$.
This relationship holds in many cases of interest, e.g.:

\begin{itemize}
\item  The Hamiltonian $H$ given by the eq.(\ref{5-2}) of the next section,
where $V^{(2)}(\omega ,\omega ^{\prime })=0$.

\item  The Hamiltonian $H$ given by the eq.(1) of paper \cite{LCIB}, where
the Hamiltonian of the proper system is $\Omega |1\rangle \langle 1|$, the
Hamiltonian of the environment is $\int_{0}^{\infty }\omega |\omega \rangle
\langle \omega ^{\prime }|d\omega $, and 
\begin{equation}
\int_{0}^{\infty }\int_{0}^{\infty }V^{(1)}(\omega ,\omega ^{\prime
})\,d\omega d\omega ^{\prime }\rightarrow \int_{0}^{\infty }g(\omega
)(|1\rangle \langle \omega |+|\omega \rangle \langle 1|)\,d\omega ,\qquad
V^{(2)}(\omega ,\omega ^{\prime })\rightarrow 0  \label{4-4}
\end{equation}

\item  The Hamiltonian $H$ given by the eq.(2.1) of paper \cite{ACGI}, where
the Hamiltonian of the proper system is $\Omega a^{\dagger }a$, the
Hamiltonian of the environment is $\int_{0}^{\infty }\omega b_{{\bf k}%
}^{\dagger }b_{{\bf k}}d{\bf k}${\bf ,} and 
\begin{equation}
\int_{0}^{\infty }\int_{0}^{\infty }V^{(1)}(\omega ,\omega ^{\prime
})\,d\omega d\omega ^{\prime }\rightarrow \int_{0}^{\infty }g(k)(a^{\dagger
}b_{{\bf k}}+b_{{\bf k}}^{\dagger }a)\,d{\bf k},\qquad V^{(2)}(\omega
,\omega ^{\prime })\rightarrow 0  \label{4-5}
\end{equation}
\end{itemize}

{\bf a) First interaction: }Since $V^{(1)}\gg V^{(2)}$, in a first step we
can neglect $V^{(2)}$ and consider the Hamiltonian (see \cite{DT})

\begin{equation}
H^{(1)}=H_{0}+V_{1}=\int_{0}^{\infty }\omega |\omega \rangle \langle \omega
|\,d\omega +\int_{0}^{\infty }\int_{0}^{\infty }V^{(1)}(\omega ,\omega
^{\prime })|\omega \rangle \langle \omega ^{\prime }|\,d\omega d\omega
^{\prime }  \label{4-6}
\end{equation}
The eigenbasis $\{|\omega \rangle _{(1)}^{+}\}$ of $H^{(1)}$ is obtained as 
\begin{equation}
|\omega \rangle _{(1)}^{+}=|\omega \rangle +\frac{1}{\omega +i0-H^{(1)}}%
V^{(1)}|\omega \rangle  \label{4-7}
\end{equation}
If we pre-multiply by $\langle \varphi |$: 
\begin{equation}
\langle \varphi |\omega \rangle _{(1)}^{+}=\langle \varphi |\omega \rangle
+\langle \varphi |\frac{1}{\omega +i0-H^{(1)}}V^{(1)}|\omega \rangle
\label{4-8}
\end{equation}
we can compute the analytical continuation of eq.(\ref{4-8}) in the lower
half-plane: 
\begin{equation}
\langle \varphi |z\rangle _{(1)}^{+}=\langle \varphi |z\rangle +\langle
\varphi |\frac{1}{z-H^{(1)}}V^{(1)}|z\rangle  \label{4-9}
\end{equation}
and obtain the complex poles of such an analytical continuation. An
analogous procedure can be followed to obtain the complex poles of the
initial condition $\rho _{0}$. On this basis, the decoherence time $%
t_{D}^{(1)}$ resulting from the first interaction turns out to be $%
t_{D}^{(1)}=\hbar /\gamma ^{(1)}$, where $\gamma ^{(1)}$ is the imaginary
part of the pole closer to the real axis (see \cite{DT}). Therefore, for
times $t\gg t_{D}^{(1)}$, the state $\rho (t)$ can be considered nearly
diagonal for all practical purposes.\footnote{%
For a complete example of this point, see in paper \cite{Alimañas} the
exhaustive analysis of the way in which the Friedrich model reaches
equilibrium in the discrete and in the continuous case. In particular, see
figures 3-7 and the computation of the pole in eq.(39). Nevertheless, in
that paper there is no reference to decoherence because at that time the
present analysis of the problem was not yet developed.}

{\bf b) Second interaction: }However, the state $\rho (t)$ has not
completely decohered yet, because the interaction $V^{(2)}$, even if very
small, is still present. Then, after the first period where $V^{(1)}$ is
dominant, for times $t\gg t_{D}^{(1)}$, $V^{(2)}$ becomes relevant; in this
situation, the total Hamiltonian can be written as 
\begin{equation}
H=H^{(1)}+V_{2}=\int_{0}^{\infty }\omega |\omega \rangle _{(1)}^{+}\langle
\omega |_{(1)}^{+}\,d\omega +\int_{0}^{\infty }\int_{0}^{\infty
}V^{(2)\prime }(\omega ,\omega ^{\prime })\,|\omega \rangle
_{(1)}^{+}\langle \omega ^{\prime }|_{(1)}^{+}\,d\omega d\omega ^{\prime }
\label{4-10}
\end{equation}
where $V^{(2)\prime }(\omega ,\omega ^{\prime })$ is $V^{(2)}(\omega ,\omega
^{\prime })$ in the new basis $\{|\omega \rangle _{(1)}^{+}\}$. Now, the
eigenbasis $\{|\omega \rangle ^{+}\}$ of $H$ is obtained as 
\[
|\omega \rangle ^{+}=|\omega \rangle _{(1)}^{+}+\frac{1}{\omega +i0-H}%
V^{(2)}|\omega \rangle _{(1)}^{+} 
\]
By repeating the procedure applied in the case of the first interaction, we
can compute the decoherence time $t_{D}^{(2)}$ obtained by taking into
account the second interaction, which results $t_{D}^{(2)}=\hbar /\gamma
^{(2)}$, where $\gamma ^{(2)}$ is again the imaginary part of the pole
closer to the real axis (see \cite{DT}). For times $t\gg t_{D}^{(2)}$, the
state $\rho (t)$ can be considered completely diagonal.

{\bf c) Estimating and comparing results: }As proved in paper \cite{DT}, $%
\gamma ^{(1)}$ and $\gamma ^{(2)}$ are proportional to the corresponding
interactions; therefore, the decoherence times $t_{D}^{(1)}$ and $%
t_{D}^{(2)} $ are proportional to the inverse of the interaction. If $%
V^{(1)} $ is a macroscopic interaction, then $t_{D}^{(1)}\approx 10^{-39}s$;
in turn, if $V^{(2)}$ is a microscopic interaction, the decoherence time $%
t_{D}^{(1)}$ may be of the order of $10^{-15}s$ (see details in \cite{DT}).
As expected, $t_{D}^{(1)}\ll t_{D}^{(2)}$.

When these general results are applied to our case, where a closed system $U$
is partitioned into a proper system $S$ and an environment $E$, they acquire
a new meaning. In fact, since $V^{(1)}$ represents the interaction between $%
S $ and $E$, $t_{D}^{(1)}$ turns out to be the decoherence time $t_{DS}$ of
the system $S$ in interaction with its environment $E$. In turn, since $%
V^{(2)}$ represents the interaction of the parts of the environment $E$
among themselves, when it is included in the total Hamiltonian (see eq.(\ref
{4-10})), the time $t_{D}^{(2)}$ turns out to be the decoherence time $%
t_{DU} $ of the whole composite system $U=S\cup E$. As expected, $t_{DS}\ll
t_{DU}$: in general, the time that a whole system needs to decohere is much
longer than the decoherence time of a small subsystem strongly coupled with
the rest of the degrees of freedom.

Summing up, from this general perspective we can describe a two-times
process, with an extremely short decoherence time for the open subsystem $S$%
, and a long (even infinite, if $V^{(2)}=0$) decoherence time for the whole
closed system $U$.

\section{A well-known model}

In this section we will apply the general theoretical framework just
presented to an example extensively treated in the literature on
decoherence. This task will allow us to draw certain conceptual conclusions
that may remain obscure when the model is studied exclusively by means of
numerical techniques.

Let us consider a system $S$ consisting in a single spin-1/2 particle $S_{0}$
(with Hilbert space ${\cal H}_{0}$), and its environment $E$ composed by a
collection of $N$ spin-1/2 particles $S_{i}$ (with Hilbert spaces ${\cal H}%
_{i}$). If the free Hamiltonians of the system and the environment are
assumed to be zero: 
\begin{equation}
H_{S}=H_{E}=0  \label{5-1}
\end{equation}
the total Hamiltonian $H=H_{S}+H_{E}+H_{SE}$ of the composite system $%
U=S\cup E$ reads (see \cite{Zurek-1} \cite{Max}) 
\begin{equation}
H=H_{SE}=\frac{1}{2}(|0\rangle \langle 0|-|1\rangle \langle
1|)\sum_{i=1}^{N}g_{i}(|\uparrow _{i}\rangle \langle \uparrow
_{i}|-|\downarrow _{i}\rangle \langle \downarrow _{i}|)\bigotimes_{j\neq
i}^{N}I_{j}  \label{5-2}
\end{equation}

Let us consider a pure initial state of the system $U$: 
\begin{equation}
|\psi _{0}\rangle =(a\,|0\rangle +b\,|1\rangle )\bigotimes_{i=1}^{N}(\alpha
_{i}|\uparrow _{i}\rangle +\beta _{i}|\downarrow _{i}\rangle )  \label{5-3}
\end{equation}
where $\alpha _{i}$ and $\beta _{i}$ are aleatory coefficients such that 
%TCIMACRO{\TEXTsymbol{\vert}}
%BeginExpansion
\mbox{$\vert$}%
%EndExpansion
$\alpha _{i}|^{2}+|\beta _{i}|^{2}=1$. The state $|\psi _{0}\rangle $
evolves as 
\begin{equation}
|\psi (t)\rangle =a\,|0\rangle \,|{\cal E}_{0}(t)\rangle +b\,|1\rangle \,|%
{\cal E}_{1}(t)\rangle  \label{5-4}
\end{equation}
where 
\begin{equation}
|{\cal E}_{0}(t)\rangle =|{\cal E}_{1}(-t)\rangle
=\bigotimes_{i=1}^{N}(\alpha _{i}\,e^{ig_{i}t/2}\,|\uparrow _{i}\rangle
+\beta _{i}\,e^{-ig_{i}t/2}\,|\downarrow _{i}\rangle )  \label{5-5}
\end{equation}
The density matrix corresponding to this state will be $\rho (t)=|\psi
(t)\rangle \langle \psi (t)|$.

Now we will analyze this model from the perspective given by steps 1 to 3 of
Section II.

1. {\bf First step:} The relevant observables $O_{R}\in {\cal O}$ for this
case will be of the form (see eq.(18) of \cite{Max})

\begin{equation}
O_{R}=(s_{00}|0\rangle \langle 0|+s_{01}|0\rangle \langle 1|+s_{10}|1\rangle
\langle 0|+s_{11}|1\rangle \langle 1|)\bigotimes_{i=1}^{N}(\epsilon
_{\uparrow \uparrow }^{(i)}|\uparrow _{i}\rangle \langle \uparrow
_{i}|+\epsilon _{\uparrow \downarrow }^{(i)}|\uparrow _{i}\rangle \langle
\downarrow _{i}|+\epsilon _{\downarrow \uparrow }^{(i)}|\downarrow
_{i}\rangle \langle \uparrow _{i}|+\epsilon _{\downarrow \downarrow
}^{(i)}|\downarrow _{i}\rangle \langle \downarrow _{i}|)  \label{5-6}
\end{equation}
where $s_{00}$, $s_{11}$, $\epsilon _{\uparrow \uparrow }^{(i)}$, $\epsilon
_{\downarrow \downarrow }^{(i)}$ are real numbers, and $s_{01}=s_{10}^{*}$, $%
\epsilon _{\uparrow \downarrow }^{(i)}=\epsilon _{\downarrow \uparrow
}^{(i)*}$ are complex numbers.

2. {\bf Second step:} The expectation value of any observable $O_{R}\in 
{\cal O}$ in the state $\psi (t)$ can be computed as

\begin{equation}
\langle O_{R}\rangle _{\psi (t)}=(|a|^{2}s_{00}+|b|^{2}s_{11})\,\Gamma
_{0}(t)+2%
%TCIMACRO{\func{Re} }
%BeginExpansion
\mathop{\rm Re}%
%EndExpansion
\,[ab^{*}\,s_{10}\,\Gamma _{1}(t)]  \label{5-7}
\end{equation}
where 
\begin{equation}
\Gamma _{0}(t)=\prod_{i=1}^{N}\left[ |\alpha _{i}|^{2}\epsilon _{\uparrow
\uparrow }^{(i)}+|\beta _{i}|^{2}\epsilon _{\downarrow \downarrow
}^{(i)}+\alpha _{i}{}^{*}\beta _{i}\epsilon _{\uparrow \downarrow
}^{(i)}e^{-ig_{i}t}+(\alpha _{i}{}^{*}\beta _{i}\epsilon _{\uparrow
\downarrow }^{(i)})^{*}e^{ig_{i}t}\right]  \label{5-8}
\end{equation}
\begin{equation}
\Gamma _{1}(t)=\prod_{i=1}^{N}\left[ |\alpha _{i}|^{2}\epsilon _{\uparrow
\uparrow }^{(i)}e^{ig_{i}t}+|\beta _{i}|^{2}\epsilon _{\downarrow \downarrow
}^{(i)}e^{-ig_{i}t}+\alpha _{i}{}^{*}\beta _{i}\epsilon _{\uparrow
\downarrow }^{(i)}+(\alpha _{i}{}^{*}\beta _{i}\epsilon _{\uparrow
\downarrow }^{(i)})^{*}\right]  \label{5-9}
\end{equation}
At this point, we will consider two particular cases:

{\bf Case (a)}: When $\epsilon _{\uparrow \uparrow }^{(i)}=\epsilon
_{\downarrow \downarrow }^{(i)}=1$ and $\epsilon _{\uparrow \downarrow
}^{(i)}=0$, the model is a typical example for EID, where the relevant
observables are only those corresponding to the proper system $S=S_{0}$
strongly coupled with its environment. In fact, these relevant observables $%
O_{R_{0}}\in {\cal O}_{0}$ read (see eq.(\ref{2-10})) 
\begin{equation}
\text{ }O_{R_{0}}=\left( \sum_{s,s^{\prime }=0,1}s_{ss^{\prime }}|s\rangle
\langle s^{\prime }|\right)
\bigotimes_{i=1}^{N}I_{i}=O_{S_{0}}\bigotimes_{i=1}^{N}I_{i}  \label{5-10}
\end{equation}
and their expectation values in the state $\psi (t)$ of $U$ result 
\begin{equation}
\langle O_{R_{0}}\rangle _{\psi (t)}=|a|^{2}\,s_{00}+|b|^{2}\,s_{11}+%
%TCIMACRO{\func{Re} }
%BeginExpansion
\mathop{\rm Re}%
%EndExpansion
[ab^{*}\,s_{10}\,r(t)]  \label{5-11}
\end{equation}
where 
\begin{equation}
r(t)=\langle {\cal E}_{1}(t)\rangle |{\cal E}_{0}(t)\rangle  \label{5-12}
\end{equation}
and 
\begin{equation}
|r(t)|^{2}=\prod_{i=1}^{N}(|\alpha _{i}|^{4}+|\beta _{i}|^{4}+2|\alpha
_{i}|^{2}|\beta _{i}|^{2}\cos 2g_{i}t)  \label{5-13}
\end{equation}

{\bf Case (b)}: However, we can also decide to ''observe'' just one particle 
$S_{j}$ of the environment, that is, to consider the observables
corresponding to $S_{j}$ as the relevant ones. These relevant observables $%
O_{R_{j}}\in {\cal O}_{j}$ read (see eq.(\ref{2-10}))

\begin{equation}
O_{R_{j}}=I_{0}\otimes O_{S_{j}}\bigotimes_{i\neq j}I_{i}  \label{5-14}
\end{equation}
where 
\begin{equation}
O_{S_{j}}=\epsilon _{\uparrow \uparrow }^{(j)}\,|\uparrow _{j}\rangle
\langle \uparrow _{j}|+\epsilon _{\downarrow \downarrow }^{(j)}\,|\downarrow
_{j}\rangle \langle \downarrow _{j}|+\epsilon _{\downarrow \uparrow
}^{(j)}\,|\downarrow _{j}\rangle \langle \uparrow _{j}|+\epsilon _{\uparrow
\downarrow }^{(j)}\,|\uparrow _{j}\rangle \langle \downarrow _{j}|
\label{5-15}
\end{equation}
and $\epsilon _{\uparrow \uparrow }^{(j)}$, $\epsilon _{\downarrow
\downarrow }^{(j)}$, $\epsilon _{\downarrow \uparrow }^{(j)}$ are now
generic. In this case, the expectation value of $O_{R_{j}}$ in the state $%
\psi (t)$ reads 
\begin{eqnarray}
\langle O_{R_{j}}\rangle _{\psi (t)}=\langle \psi (t)|O_{R_{j}}|\psi
(t)\rangle = &&|a|^{2}(|\alpha _{j}|^{2}\epsilon _{\uparrow \uparrow
}^{(j)}+|\beta _{j}|^{2}\epsilon _{\downarrow \downarrow }^{(j)}+\alpha
_{j}\beta _{j}^{*}\epsilon _{\uparrow \downarrow }^{(j)}e^{-ig_{j}t}+\alpha
_{j}^{*}\beta _{j}\epsilon _{\downarrow \uparrow }^{(j)}e^{ig_{j}t})+ 
\nonumber \\
&&+|b|^{2}(|\alpha _{j}|^{2}\epsilon _{\uparrow \uparrow }^{(j)}+|\beta
_{j}|^{2}\epsilon _{\downarrow \downarrow }^{(j)}+\alpha _{j}\beta
_{j}^{*}\epsilon _{\uparrow \downarrow }^{(j)}e^{ig_{j}t}+\alpha
_{j}^{*}\beta _{j}\epsilon _{\downarrow \uparrow }^{(j)}e^{-ig_{j}t})
\label{5-16}
\end{eqnarray}

3. {\bf Third step:} The time evolution of the expectation values of the
relevant observables can be computed in both cases:

{\bf Case (a)}: Since 
\begin{eqnarray}
\max_{t}(|\alpha _{i}|^{4}+|\beta _{i}|^{4}+2|\alpha _{i}|^{2}|\beta
_{i}|^{2}\cos 2g_{i}t) &=&1  \label{5-17} \\
\min_{t}(|\alpha _{i}|^{4}+|\beta _{i}|^{4}+2|\alpha _{i}|^{2}|\beta
_{i}|^{2}\cos 2g_{i}t) &=&(2|\alpha _{i}|^{2}-1)^{2}  \label{5-18}
\end{eqnarray}
$(|\alpha _{i}|^{4}+|\beta _{i}|^{4}+2|\alpha _{i}|^{2}|\beta _{i}|^{2}\cos
2g_{i}t)$ is an aleatory number that, if $t\neq 0$, fluctuates between $1$
and $(2|\alpha _{i}|^{2}-1)^{2}$. Then, from eq.(\ref{5-13}) we can conclude
that, for $N\rightarrow \infty $, 
\begin{equation}
\lim_{t\rightarrow \infty }r(t)=0  \label{5-19}
\end{equation}
In turn, from eq.(\ref{5-11}) we obtain the limit 
\begin{equation}
\lim_{t\rightarrow \infty }\langle O_{R_{0}}\rangle _{\psi
(t)}=|a|^{2}\,s_{00}+|b|^{2}\,s_{11}=\langle O_{R_{0}}\rangle _{\rho _{*}}
\label{5-20}
\end{equation}
where $\rho _{*}$ is the final diagonal state of $U$. This result can also
be expressed as a weak limit for $\rho (t)=|\psi (t)\rangle \langle \psi
(t)| $ (see eq.(\ref{2-17})): 
\begin{equation}
W-\lim_{t\rightarrow \infty }\rho (t)=\rho _{*}  \label{5-21}
\end{equation}
If we now consider the reduced density operator $\rho _{S_{0}}$
corresponding to the system $S=S_{0}$, eq.(\ref{5-20}) can be expressed as
(see eq.(\ref{2-15})) 
\begin{equation}
\lim_{t\rightarrow \infty }\langle O_{R_{0}}\rangle _{\rho (t)}=\langle
O_{R_{0}}\rangle _{\rho _{*}}=\lim_{t\rightarrow \infty }\langle
O_{S_{0}}\rangle _{\rho _{S_{0}}(t)}=\langle O_{S_{0}}\rangle _{\rho
_{S_{0*}}}  \label{5-22}
\end{equation}
where

\begin{equation}
\rho _{S_{0}*}=\left( 
\begin{array}{ll}
|a|^{2} & 0 \\ 
0 & |b|^{2}
\end{array}
\right)  \label{5-23}
\end{equation}
This result shows that, as expected, the system $S=$ $S_{0}$ in interaction
with the environment $E$ decoheres in the eigenbasis of $\rho _{S_{0}*}$.

{\bf Case (b)}: If we decide to ''observe'' the particle $S_{j}$ of the
environment, we have to consider the evolution of the expectation value of
the corresponding relevant observables $O_{R_{j}}\in {\cal O}_{j}$. Eq.(\ref
{5-16}) shows that $\langle O_{R_{j}}\rangle _{\psi (t)}$ just oscillates
and, therefore, it has no limit for $t\rightarrow \infty $. As a
consequence, a generic particle $S_{j}$ of the environment does not
decohere. This result is completely foreseeable from a physical point of
view: to the extent that the particles $S_{i}$ of the environment $%
E=\bigcup_{i}S_{i}$ are uncoupled to each other, they freely evolve;
therefore, the environment composed by these freely evolving particles is
unable to reach a final decohered state.

The results just obtained point to the fact that we can gain a better
understanding of the behavior of a closed system by studying the behavior of
its subsystems. In our case, the behavior of the whole system $U$ can be
completely described by analyzing only the observables corresponding to
cases (a) and (b). In fact, the total Hamiltonian $H$ of eq.(\ref{5-2}) is
not symmetric with respect to the particle $S_{0}$ and the generic particle $%
S_{j}$: whereas $S_{0}$ is coupled to all the $S_{i}$ of the environment,
the $S_{j}$ do not interact among themselves and they are only coupled to $%
S_{0}$ (and this coupling vanishes when $N\rightarrow \infty $). Therefore, $%
S=S_{0}$ decoheres in interaction with the environment $E$ in a finite
decoherence time $t_{DS}$, but the environment $E$ does not decohere, that
is, it has an infinite decoherence time resulting from the trivial
interaction among its parts.

Now, this result allows us to conceptually infer the behavior of the whole
composite system. As we conclude at the end of Subsection III.A, the
decoherence of the whole composite system imposes a physical condition
strong enough to imply the decoherence of any of its subsystems. Therefore,
if any subsystem of a closed composite system does not decohere, then we can
be sure that the whole composite system will neither decohere. This is
precisely the case of our model: since $S$ decoheres but $E$ does not
decohere, the composite system $U=S\cup E$ cannot decohere, that is, it has
an infinite decoherence time, $t_{DS}\ll t_{DU}=\infty $.\footnote{%
When this well-known model is understood from our general perspective, the
criticisms to SID presented in paper \cite{Max} vanish (for a detailed
criticism of \cite{Max}, see \cite{EIDSIDSynthese}). In fact, the model does
not show that the destructive interference of the off-diagonal terms is not
always efficient, as claimed, but it merely proves that the closed system
does not decohere because the Hamiltonian of the environment is trivial, and
this -physically obvious- fact is perfectly explained in the context of the
SID approach.}

Summing up, the arguments presented in this section show that, when this
well-known model is analyzed in the context of our theoretical framework,
the results obtained in the special case can be viewed from a new general
perspective. In particular, certain results computed by means of numerical
techniques and which may seem puzzling when considered in isolation, turn
out to be necessary conceptual consequences of the full understanding of the
physical phenomenon.

\section{Conclusions}

In this paper we have presented a common theoretical framework that
encompasses both EID and SID, and probably other decoherence approaches.
When it is accepted that the formalisms of decoherence for open and closed
systems cooperate in the understanding of the same physical phenomenon, the
results obtained by means of the different theories of decoherence can be
retained as relevant acquisitions: for instance, the large amount of
experimental confirmations of EID (see \cite{Ex}), or the complete
description of the classical limit of quantum mechanics \cite{CL} and the
study of the role of complexity in decoherence \cite{QC} in the case of SID,
or the extremely short decoherence times computed by EID and SID \cite{DT}.

In turn, from the new general perspective just proposed, the difficulties
that the EID approach has to face (see Introduction) are not as serious as
originally supposed. In fact,

\begin{itemize}
\item  It can be explained that closed systems do decohere; moreover, the
decoherence time of a closed system can be computed.

\item  The serious conceptual problem of deciding where to place the cut
between ''the'' proper system $S$ and ''the'' environment $E$ in a composite
system $U=S\cup E$ is dissolved, because the closed system $U$ may be
partitioned in several different ways: no one of them is the ''true'' or
''correct'' partition. When we want to understand the behavior of the whole
system $U$, we have to select certain partitions to study the resulting
subsystems: the choice of the relevant partitions is guided by the
inspection of the total Hamiltonian.

\item  The final pointer bases in which the open and the closed systems
decohere are well defined in the corresponding weak limits, with complete
generality.
\end{itemize}

Finally, it is interesting to note that, since from this general perspective
decoherence is conceived as a particular case of irreversible phenomena, the
conclusions drawn in this paper may serve to illuminate some traditional
issues of the problem of irreversibility. For instance, the old question
about the criterion for selecting the relevant macroscopic variables of a
system may receive a simple answer: we can choose different sets of
macroscopic variables, but we have to study the behavior of more than one
set if we want to reach the understanding of the behavior of the whole
irreversibly evolving system.

\section{Acknowledgments}

We are very grateful to Roland Omn\'{e}s and Maximilian Schlosshauer for
many comments and criticisms. This research was partially supported by
grants of the University of Buenos Aires, the CONICET and the FONCYT of
Argentina.

\appendix 

\section{Coarse-graining and projection}

As it is well-known, a coarse-graining amounts to a projection whose action
is to eliminate some components of the state vector corresponding to the
finer description. If this idea is generalized, coarse-graining can be
conceived as a projection that reduces the number of components of a
generalized vector representing a state. In the light of this idea, in
Section II we have argued that, for any observable $O_{R}$ belonging to the
space ${\cal O}$ of relevant observables, the expectation value of $O_{R}$
in the state $\rho (t)$ can be expressed in terms of a coarse-grained state $%
\rho _{G}(t)$ such that $\langle O_{R}\rangle _{\rho (t)}=\langle
O_{R}\rangle _{\rho _{G}(t)}$. In this Appendix, we will prove that: (i) the 
$\rho _{G}(t)$ so defined is the result of the projection of the state $\rho
(t)$ onto the space ${\cal O}$ of relevant observables, (ii) the final state 
$\rho _{G*}$ of $\rho _{G}(t)$ is the result of the projection of the final
state $\rho _{*}$ of $\rho (t)$ onto ${\cal O}$, and (iii) when ${\cal O}$
has a finite number of dimensions, for $t\rightarrow \infty $, $\rho _{G}(t)$
tends to $\rho _{G*}$ not only in a weak sense but also in a strong sense.

(i).- Let us use the notation $\langle O\rangle _{\rho }=(\rho |O)$. Let the
basis of ${\cal O}${\it \ }be{\it \ }$\{|O_{Rj})\}$, and let us define a
projector 
\begin{equation}
\pi =\sum_{i=1}^{n}|O_{Ri})(\rho _{i}|,\quad (\rho _{i}|O_{Rj})=\delta _{ij}
\label{A-1}
\end{equation}
where the $(\rho _{i}|$ are functionals defined by $(\rho
_{i}|O_{Rj})=\delta _{ij}$.\footnote{%
If we are working in a finite dimensional space ${\cal O}$, we can choose $%
|O_{Rj})=|\alpha \rangle \langle \beta |$ with $\langle \alpha |\beta
\rangle =\delta _{\alpha \beta }$ and, then, $(\rho _{i}|=|\alpha \rangle
\langle \beta |$.} Obviously, $\pi ^{2}=\pi $. Then, we define 
\begin{equation}
(\rho _{G}(t)|=(\rho (t)|\pi   \label{A-2}
\end{equation}
Now, 
\begin{equation}
(\rho _{G}(t)|O_{Rj})=(\rho (t)|\pi |O_{Rj})=(\rho
(t)|\sum_{i=1}^{n}|O_{Ri})(\rho _{i}|O_{Rj})=(\rho
(t)|\sum_{i=1}^{n}|O_{Ri})\delta _{ij}=(\rho (t)|O_{Rj})  \label{A-3}
\end{equation}
Then, making linear combinations, we obtain 
\begin{equation}
(\rho (t)|O_{R})=(\rho _{G}(t)|O_{R})\text{,\quad if }|O_{R})\in {\cal O}
\label{A-4}
\end{equation}

(ii).- But now 
\begin{equation}
(\rho _{G}(t)|O_{Rj})=(\rho (t)|\pi |O_{Rj})=\sum_{i=1}^{n}(\rho
(t)|O_{Ri})(\rho _{i}|O_{Rj})  \label{A-5}
\end{equation}
So, using eq.(\ref{2-0}), we obtain 
\begin{eqnarray}
\lim_{t\rightarrow \infty }(\rho _{G}(t)|O_{Rj}) &=&\lim_{t\rightarrow
\infty }(\rho (t)|\pi |O_{Rj})=\lim_{t\rightarrow \infty
}\sum_{i=1}^{n}(\rho (t)|O_{Ri})(\rho _{i}|O_{Rj})=  \nonumber \\
&=&\sum_{i=1}^{n}(\rho _{*}|O_{Ri})(\rho _{i}|O_{Rj})=(\rho _{*}|\pi
|O_{Rj})=(\rho _{G*}|O_{Rj})  \label{A-6}
\end{eqnarray}
where we have defined 
\begin{equation}
(\rho _{G*}|=(\rho _{*}|\pi   \label{A-7}
\end{equation}
And from eq.(\ref{A-6}) we obtain 
\begin{equation}
W-\lim_{t\rightarrow \infty }(\rho _{G}(t)|=(\rho _{G*}|  \label{A-8}
\end{equation}

(iii).- In the special case that ${\cal O}$ as a finite number $n$ of
dimensions, we can compute the finite number of coordinates of $(\rho
_{G}(t)|$ and $(\rho _{G*}|$:\footnote{%
In fact, with the notation of the last footnote, $\rho _{Gi}(t)=(\rho
_{G}(t)|O_{Ri})=Tr(\rho _{G}(t)O_{Ri})=\langle \beta |\rho _{G}(t)|\alpha
\rangle $ $=\rho _{G\beta \alpha }(t)$.} 
\begin{equation}
\rho _{Gi}(t)=(\rho _{G}(t)|O_{Ri}),\text{ \quad }\rho _{G*i}=(\rho
_{G*}|O_{Ri})  \label{A-9}
\end{equation}
So, from eq.(\ref{A-6}), 
\begin{equation}
\lim_{t\rightarrow \infty }\rho _{Gi}(t)=\rho _{G*i}  \label{A-10}
\end{equation}
These are simple limits of the coordinates. But, since ${\cal O}$ has finite
dimension, we obtain the strong limit 
\begin{equation}
S-\lim_{t\rightarrow \infty }(\rho _{G}(t)|=(\rho _{G*}|  \label{A-11}
\end{equation}

\end{document}